\documentclass[10pt,superscriptaddress]{article}

\usepackage[letterpaper, margin=.9in]{geometry}

\usepackage{amsmath,amsfonts,amsthm,amssymb,color}
\usepackage[hidelinks]{hyperref}
\usepackage{fancybox}
\usepackage{physics}
\usepackage{framed}
\usepackage{algpseudocode}
\usepackage{comment}
\usepackage{mathrsfs}
\usepackage{hyperref}
\usepackage{cleveref}
\usepackage{subcaption}
\usepackage{xcolor}
\usepackage{authblk}

\usepackage{enumitem}
\usepackage{color,soul}
\usepackage{multicol}
\usepackage{graphicx}

\usepackage[style=ieee, citestyle=numeric-comp,sorting=none]{biblatex} 
\usepackage{footnote}
 \usepackage[ruled,vlined,linesnumbered]{algorithm2e}

\theoremstyle{definition}

\DeclareMathOperator*{\argmin}{argmin}

\begin{document}

\title{Calculating the expected value function of a two-stage stochastic optimization program with a quantum algorithm}

%
\author[1]{Caleb Rotello\thanks{caleb.rotello@nrel.gov}}
\author[1]{Peter Graf}
\author[1]{Matthew Reynolds}
\author[2]{Eric B. Jones}
\author[1]{Cody James Winkleblack}
\author[1]{Wesley Jones}
\affil[1]{Computational Science Center, National Renewable Energy Laboratory, Golden CO, 80401}
\affil[2]{Infleqtion, Louisville CO, 80027}

\maketitle

\begin{abstract}
Two-stage stochastic programming is a problem formulation for decision-making under uncertainty. In the first stage, the actor makes a best ``here and now'' decision in the presence of uncertain quantities that will be resolved in the future, represented in the objective function as the \textit{expected value function}. This function is a multi-dimensional integral of the second stage optimization problem, which must be solved over all possible future scenarios. This work uses a quantum algorithm to estimate the expected value function with a polynomial speedup. Our algorithm gains its advantage through the two following observations. First, by encoding the probability distribution as a quantum wavefunction in an auxilliary register, and using this register as control logic for a phase-separation unitary, Digitized Quantum Annealing (DQA) can converge to the minimium of each scenario in the random variable in parallel. Second, Quantum Amplitude Estimation (QAE) on DQA can calculate the expected value of this per-scenario optimized wavefunction, producing an estimate for the expected value function. Quantum optimization is theorized to have a polynomial speedup for combinatorial optimization problems, and estimation error from QAE is known to converge inverse-linear in the number of samples (as opposed to the best case inverse of a square root in classical Monte Carlo). Therefore, assuming the probability distribution wavefunction can be prepared efficiently, we conclude our method has a polynomial speedup (of varying degree, depending on the optimization problem) over classical methods for estimating the expected value function. We conclude by demonstrating this algorithm on a stochastic programming problem inspired by operating the power grid under weather uncertainty.

\end{abstract}


\section{Introduction}\label{sec:intro}
Two-stage stochastic programming is a problem formulation for decision-making under uncertainty, where an actor makes ``here and now" decisions in the presence of uncertain quantities that will be resolved at a later time; important problems like energy planning~\cite{haneveld2000}, traffic routing~\cite{laporte1992}, and industrial planning~\cite{dempster1981} can be formed as these programs. Decisions made after the realization of unknown quantities, sometimes known as recourse or second-stage decisions, are also included in the two-stage problem formulation. When optimizing under uncertainty, one chooses how to account for uncertainty in the problem formulation: For two-stage stochastic programming, uncertainty is modeled using random variables over quantities that are only revealed in the second stage, and enters the objective function using an expectation over suboptimization problems with respect to these variables. There are multiple formulations outside the scope of this paper, including robust formulations and chance constraints; we refer the interested reader to~\cite{shapiro2021lectures} for a review of these formulations.  

Heuristically, one can think of using an expectation as a mechanism for making optimal decisions ``on average" (as compared to preparing for the worst-case scenario, as is the case in robust problem formulations). More formally, a general problem formulation for two-stage stochastic programming is:
\begin{align}
\min_{x} \quad & o\left(x\right) := f\left(x\right) + \mathbb{E}_\xi\left[Q\left(x,\xi\right)\right]\, \nonumber \\ 
\text{s. t. } \quad & Ax \leq b
\end{align}
where $x\in \mathbb{R}^{n_x}$ is the first-stage decision vector and $n_x$ is the number of first-stage decision variables; $f:\mathbb{R}^{n_x} \rightarrow \mathbb{R}$ is a function depending only on the first-stage decision $x$. If $\xi$ is a discrete random variable over which the expectation above is calculated\footnote{$\xi: \Omega \rightarrow \Xi$ and $\Xi$ is a measurable set $\Xi \subseteq \mathbb{R}^{n_\xi}$}, then  $\xi_\omega \in \mathbb{R}^{n_\xi}$ is a realization of the random variable as a vector with $n_\xi$ components (i.e., a sample of the random varaible) with probability mass function $p(\omega)$, where $\omega \in \Omega$ is an element of a discrete sample space $\Omega$. In the example of a single coin flip, we can write the discrete sample space as $\Omega = \{H,\,T\}$, and the random variable maps these values to real numbers $\xi_H = -1$ and $\xi_T = 1$. This is useful, because it allows us to actually perform a calculation like $p(T)\xi_T+p(H)\xi_H=-1/2+1/2=0$, which demonstrates that the coin flip is unbiased. Let $Ax\leq b$ represent an arbitrary constraint on $x$. Finally, $Q\left(x,\xi\right)$ is the second-stage cost function defined by\footnote{We define $Q$ using the random variable $\xi$; to evaluate the function, we use a specific realization $\xi_\omega$.}
\begin{align}
Q\left(x,\xi\right) =  \min_{y} \quad & q\left(x,y,\xi\right)  \nonumber \\
\text{s. t.} \quad & T(\xi)x + W(\xi)y = h(\xi), 
\label{eq:QQQ}
\end{align}
where $y\in \mathbb{R}^{n_y}$ are the second-stage variables and $n_y$ is the number of second-stage decision variables; the second-stage cost function is $q: \mathbb{R}^{n_x+n_y+n_\xi} \rightarrow \mathbb{R}$; and $T(\xi)x + W(\xi)y = h(\xi)$ represents an arbitrary constraint on $x$ and $y$ depending on the random variable $\xi$. For this work, we only consider programs where $q(x,y,\xi)$ can be bounded as $q_l \leq q(\bullet,\bullet,\bullet) \leq q_u$ where $q_l$ and $q_u$ can be found (or at least assumed). If this assumption is violated, the results still hold, it would just change the derived error bound~\cite{Montanaro_2015}. This problem formulation seeks the minimum first stage decision, $x^* = \argmin_{x} o(x)$, where the cost of a first stage decision is assessed by adding together some first stage deterministic function $f$ and the average cost of how that decision will perform in the future, which is represented by the expected value function $\mathbb{E}\left[Q(x,\xi)\right]$. Each scenario $\xi_\omega$ will dictate its own optimal second-stage decision $y_\omega ^* = \argmin_y q(x,y,\xi_\omega)$. For the remainder of the manuscript, we define the expected value function 
\begin{equation}
    \phi(x) := \mathbb{E}_\xi\left[ Q(x,\xi) \right].
\end{equation}

If $\xi$ is continuous, a sample-average is often employed to approximate the expected value function. This approximation, called the sample-average approximation (SAA), is computed using only a finite set of observations of $\xi$~\cite{glasserman2000, shapiro2007}. Because of this, most solution techniques operate under the assumption that one is optimizing over a discrete sample space. Thus for the rest of this work we assume that $\xi$ is a discrete random variable.

In general, computing the expected value function in a two-stage stochastic program is \#P-Hard~\cite{Dyer2006}, the complexity class belonging to counting problems, making simply \textit{evaluating} the cost function of a two-stage stochastic program theoretically much harder than the NP complexity class. This difficulty holds even if the second-stage optimization function $Q$ is linear or can otherwise be evaluated in polynomial time for a given $x,\,\xi_\omega$ pair; the problem gets promoted to higher complexity once the underlying functions are NP-Hard~\cite{PAPADIMITRIOU1985288}. Because even evaluating the objective function for a candidate first-stage decision $x$ is so difficult, actually solving two-stage stochastic programs is far more computationally intensive than already difficult NP optimization problems.

There are several ways this problem formulation is solved. Most commonly this is solved with the extensive form~\cite{10.3389/fceng.2020.622241, SHAPIRO2003353}, where the expected value function is written as a weighted sum $\mathbb{E}\left[Q(x,\xi)\right] \rightarrow \sum_{\omega \in \Omega} p(\omega) q(x,y_\omega, \xi_\omega)$ and the second stage decision vector $y$ is expanded to a different vector $y_\omega$ for each scenario $\xi_\omega$. These transformations allow the stochastic program to be cast as a larger deterministic optimization program, usually  with larger space and time complexity. Alternatively, solvers use nested programming techniques, which use an alternative computing technique, like neural networks in~\cite{dumouchelle2022neur2sp, Nair2018LearningFO, yoshua2020}, to estimate, guess, or otherwise simplify the expected value function for a candidate first-stage decision. A deterministic programming technique then iteratively improves this candidate first-stage decision. Our work provides a quantum computing algorithm to be used in nested solvers. This algorithm computes the expected value function $\phi(x)$ for a given first-stage choice, in what we estimate is time complexity polynomially faster and requiring exponentially less space than the extensive form. 

This paper is organized as follows. Sec.~\ref{subsec:overview} is an overview of our algorithm. We subsequently examine the details of its components, and hypothesize their advantages, in Secs.~\ref{sec:aqaoa} and~\ref{sec:qae}. Finally, in Sec.~\ref{sec:example} we implement the algorithm for an idealized, binary, version of the  ``unit commitment problem", a two-stage stochastic program that occurs frequently in power systems operation.

\subsection{Overview of our algorithm}\label{subsec:overview}
We propose a hybrid quantum-classical algorithm to solve this problem, where a classical computer solves \begin{equation}\label{eq:so_vqa}
\min_{x} \quad \tilde{o}(x) := f(x) + \tilde{\phi}(x),
\end{equation}
and the estimate $\tilde{\phi}(x)$ of the expected value function $\phi(x)$ is computed by our quantum algorithm. There are two main components of our algorithm. The first component, discussed in detail in Sec.~\ref{sec:aqaoa}, is a specific implementation of the Quantum Alternating Operator Ansatz (QAOA)~\cite{farhi2014quantum, Hadfield_2019} designed to mimic quantum annealing as in~\cite{Wurtz_2022}; we call this Digitized Quantum Annealing (DQA). DQA simulates an annealing evolution of time $T$ at discrete time-step size $dt$ using $T/dt$ alternations of a problem and mixing Hamiltonian to prepare an estimate of the ground state of the problem Hamiltonian, which corresponds to the solution of an optimization problem. We modify the ansatz to use a secondary register, which stores the random variable $\xi$ as a superposition of its scenarios, as control logic on the second-stage decision variables $y$, which allows us to optimize over all scenarios simultaneously. This creates a per-scenario optimized wavefunction $\ket{\psi_x^*}$ that encodes the optimal second-stage decision $y_\omega^*$ for each scenario $\xi_\omega$ w.r.t. a candidate first stage decision $x$, in time independent of any properties of the random variable. The second component, discussed in detail in Sec.~\ref{sec:qae}, uses Quantum Amplitude Estimation (QAE)~\cite{Brassard_2002}. QAE uses DQA as a subroutine to compute the estimate $\tilde{\phi}(x)$  from $\ket{\psi_x^*}$. QAE also involves an oracle operator that, given $\ket{\psi_x^*}$ and an ancilla qubit, rotates amplitude onto that ancilla qubit in proportion to the second-stage objective function value $q(x,y_\omega^*,\xi_\omega) $  of each pair $y_\omega^*,\,\xi_\omega$. The probability amplitude on the $\ket{1}$ state of this ancilla qubit is proportional to the expected value function. QAE  repeats DQA and the oracle a number of times, $M$, to estimate this probability amplitude, which allows us to extract the desired estimate of the expected value function.

There are two sources of systematic error in our estimate $\tilde{\phi}(x)$, one from each component of the algorithm. The first, from DQA, is the residual energy $\delta$  left after our annealing process. In the large $T$ limit, we expect this to decrease with a power law $\delta \sim O(1/T^2)$~\cite{Suzuki_2005}.  The second, from QAE, can be thought of as a sampling error; QAE has additive error that decreases with $O(1/M)$, where $M$ is the number of times QAE repeats the DQA and oracle circuit subroutines. This will allow us to derive the error formula
\begin{equation}\label{eq:error}
    \text{Pr}\left[\left|\frac{\tilde{\phi}(x) - \phi(x) - \delta}{q_u-q_l}\right| \leq \frac{\pi}{M} + \frac{\pi^2}{M^2}\right] \geq \frac{8}{\pi^2},
\end{equation}
where $q_u,q_l$ are bounds on the second-stage objective function, in Sec.~\ref{subsec:error}. This formula states that the result of our algorithm, $\tilde{\phi}(x)$, will have some precision $O(1/M)$ and some accuracy $\delta$. 

We hypothesize three sources of quantum advantage from our method: two in time and one in space. The first time speedup comes from the hypothesized polynomial time speedup given by quantum annealing and QAOA~\cite{Somma_2012, PhysRevX.10.021067, Guerreschi2019}. The second time speedup comes from the polynomially faster convergence of QAE over classical Monte Carlo sampling~\cite{Montanaro_2015,Giurgica_Tiron_2022}. As seen in Eq.~\ref{eq:error}, the errors of each component in the algorithm are additive, so we expect the quantum advantages of each component to be preserved, leading us to believe our algorithm has a polynomial advantage in time complexity. In big-O notation, we expect our time complexity to be $\tilde{O}(TM/dt)$, where the $\tilde{O}$ hides polylogarithmic factors. Interestingly, the time complexity of our algorithm does not depend on the size of the sample set $\Omega$ or structure of the random variable $\xi$, besides the time it takes to load the random variable $\xi$ as a wavefunction; this is because the annealing time only depends on the size of the second-stage decision vector, $n_y$, which we show in Sec.~\ref{subsec:pdfqaoa}, and the time required for QAE only depends on the target additive error, which we show in Sec.~\ref{subsec:error} does not depend on any properties of $\xi$ (also see~\cite{Montanaro_2015}). We also expect an exponential advantage in space complexity, because we store $\xi$ as a quantum wavefunction. In the worst case, the size of the sample space is $|\Omega|=2^{n_\xi}$, or exponential in the number of variables in a given scenario $\xi_\omega$. By storing $\xi$ as a superposition of its scenarios, we only need $\lceil\log_2|\Omega|\rceil =O(n_\xi)$ qubits, leading to an exponential advantage in space~\cite{grover2002creating}. In big-O notation, we expect our space complexity to be $O(n_x+n_y+n_\xi)$. 

\section{Computing the minima over a scenario set}\label{sec:aqaoa}
In this section we review the Digitized Quantum Annealing (DQA), implemented with a Quantum Alternating Operator Ansatz (QAOA) where the operator angles follow an annealing schedule (rather than being optimized variationally), and show how using a random variable stored in a secondary register as control logic for the cost operator in the ansatz can optimize a system in annealing time (i.e. layers in the QAOA) independent of the complexity of the random variable. First, we give an overview of the QAOA and how to implement DQA with it, then introduce our operators in the QAOA and how they use the random variable as control logic. We will do this by defining a Hamiltonian $H_Q$ such that, given a candidate first-stage decision $x$ and long enough annealing time $T$, the DQA can reliably prepare the per-scenario optimized wavefunction $\ket{\psi_x^*}$ such that $\bra{\psi_x^*}H_Q \ket{\psi_x^*} = \phi(x)$. Finally, we will give an argument about the unstructured search limit of this problem to calculate an upper bound on the annealing time, and show that this upper bound only depends on the number of second-stage variables, $n_y$. For the remainder of the manuscript we assume all variables have been binarized, so the number of qubits is equivalent to the number of variables.

\subsection{The Quantum Alternating Operator Ansatz for Digitized Quantum Annealing}\label{subsec:qaoa_overview}
Here we discuss the QAOA in general, and then show how use it to implement a DQA. Consider the optimization problem $y^*=\argmin_{y\in S} f(y)$, where $f$ is an objective function and $S$ is a set of feasible states that satisfy the problem's constraints. For this subsection, assume all $n$ qubits represent a decision variable (we relax this assumption below to incorporate $x$ and $\xi$, both of which are fixed when evaluating the expected value function $\phi(x)$). The QAOA will solve this problem by implementing the following three components~\cite{Hadfield_2019}: 
\begin{enumerate}
    \item An easy to prepare initial state that is a uniform superposition over all bitstrings ($Z$ basis vectors) that are candidate solutions. If $S$ is the set of all candidate (i.e. feasible) solutions, we prepare the initial superposition state with 
    \begin{equation}
        V_S\ket{0}^{\otimes n} = \frac{1}{\sqrt{|S|}} \sum_{y\in S} \ket{y}.
  \end{equation}
    \item A diagonal cost (or ``phase-separation'') Hamiltonian $H_C$, defined by 
\begin{equation}
        H_C\ket{y} = f(y) \ket{y},
\end{equation}
where $\ket{y}$ is a $Z$ basis vector. We apply this with the unitary $U_C(\gamma) = e^{-i \gamma H_C}$, which assigns a phase proportional to the objective function $f$ times some angle $\gamma$. This generally takes the form $H_C = \sum_j J_j Z_j + \sum_{j,k} J_{j,k}Z_jZ_k + \sum_{j,k,l}J_{j,k,l} Z_jZ_kZ_l + \cdots$.
for parameters $J_j, J_{j,k}, \cdots $ that encode a specific objective function. 
    \item A mixing Hamiltonian $H_M$ and corresponding $U_M(\gamma) = e^{-i \gamma H_M}$ that moves probability amplitude between different bitstrings in $S$. The state in item 1 should be the ground state of this Hamiltonian\footnote{If $\ket{S}\bra{S}$ is the projector onto $S$, we know by~\cite{PhysRevResearch.5.033082} that $[H_M, \ket{S}\bra{S}] = 0$}.
\end{enumerate}
We can then write the QAOA as
\begin{equation}\label{eq:qaoa}
    U(\theta,r) = \bigg[\prod_{l=1}^r U_M(\theta_{2l+1})U_C(\theta_{2l})\bigg]V_S,
\end{equation}
where $\theta$ is a list of operator angles and $r$ is the circuit depth, and apply it to a register of zeros $\ket{0}^{\otimes n}$, which creates the wavefunction
\begin{equation}
    \ket{\psi(\theta, r)} = U(\theta, r)\ket{0}^{\otimes n}.
\end{equation}
For the variational QAOA (often called the Quantum Approximate Optimization Algorithm) a classical optimizer solves $\min_{\theta,r} \bra{\psi(\theta,r)} H_C \ket{\psi(\theta,r)}$. The classical computer chooses $\theta$ and $r$, and the quantum processor repeatedly prepares $\ket{\psi(\theta,r)}$ and measures it to approximate the expectation value. If the procedure is successful, there will be a high probability of measuring $\ket{y^*}$ after iterating on $\theta,\,r$ many times. In the infinite $r$ limit, this probability converges to one by the adiabatic theorem~\cite{farhi2014quantum}.

For the quantum algorithm presented in this manuscript, we are more interested in preparing a wavefunction that has high overlap with the ground state, rather than just measuring the ground state with high probability (this is because when we perform Quantum Amplitude Estimation in Sec.~\ref{sec:qae}, our estimate will calculate $\langle H_C\rangle$, which will include residual energy resulting from the wavefunction not overlapping with the ground state). This leads us to DQA, which is a digital quantum computer implementation of quantum annealing~\cite{Wurtz_2022}. To implement DQA using the QAOA, we choose an annealing time $T$, sufficiently small Trotter step size $dt$ ~\cite{lloyd96}, and interpolating functions $a(t)$ and $b(t)$ where $a(0) = b(T) = 0$. Let the cost Hamiltonian angles be $\theta_{2l} = a(l*dt)$ and the mixing Hamiltonian angles be $\theta_{2l+1} = b(l*dt)$. The parameter $r$ in $U(\theta,r)$ is set as $r=T/dt$. 
For the rest of this work, we set the interpolating functions to be $a(t) = t/T$ and $b(t) = 1-t/T$, but retain the use of $\theta$ as a parameter to remind the reader that we can use annealing schedules other than linear interpolation~\cite{PhysRevApplied.17.044005}. The annealing time (and, by extension, circuit depth) $T$ is picked to satisfy the adiabatic condition; while this time is hard to know apriori, as it is highly problem dependent and often needs to be iterated on~\cite{Suzuki_2005,PhysRevE.58.5355,Yarkoni_2022,farhi1996analog,farhi2000quantum}, a generally safe worst case for optimization problems is $T=O(\sqrt{2^n})$, which is the annealing time for an unstructured search problem~\cite{PhysRevA.68.062312,Albash_2018} and is considered an extremal case. 

The specific implementations of $S$, $H_C$, and $H_M$ are problem dependent and depend on the presence or absence of constraints in the optimization problem. If there are no constraints in the problem, then $S$ is the full set of length $n$ bitstrings, and the mixing Hamiltonian is $H_M = \sum_j X_j$. Problem encoding becomes more difficult once constraints are introduced. Currently, there are two main ways of addressing constraints: by adding a penalty term in the cost function and/or using a ``constraint-preserving mixer''. A penalty term enforces ``soft'' constraints by re-writing the objective function as $f(y) \rightarrow f(y) + \lambda g(y)$, where the term $\lambda g(y)$ assigns an unfavorable value to bitstrings that violate the problem constraint, and is implemented as a separate phase separation Hamiltonian. To use the constraint-preserving mixer, we identify a symmetry present in the problem constraints and pick a mixing Hamiltonian that conserves this symmetry. As an example, consider the case where all bitstrings must have a certain Hamming weight. We then choose $S$ to be the set of all bitstrings with this Hamming weight, and choose the mixing Hamiltonian $H_M = \sum_{j,k} X_jX_k + Y_jY_k$, which conserves Hamming weight~\cite{He_2023}. For a more detailed discussion of these paradigms, we refer the reader to~\cite{Fuchs_2022}. Additionally, new methods that use the Zeno effect~\cite{Herman2023} and error correction~\cite{pawlak2023subspace} have also been developed to handle constraints in these problems.

In a problem with multiple constraints, it is possible to implement some of them via penalty terms and others via constraint preserving mixers. For the remainder of the manuscript, we assume that any penalty term in the second stage is included in the second-stage objective function $q$ and not explicitly written separately and that $V_S$ and $H_M$ are chosen to conserve the remaining constraints.

\subsection{Quantum Alternating Operator Ansatz including a discrete random variable}~\label{subsec:pdfqaoa}
We now discuss how to use DQA to optimize for each scenario in the discrete random variable $\xi$. This is done by reserving separate registers of optimization qubits and distribution qubits, and using the distribution register as control logic for operators acting on the optimization register in the QAOA. Specifically, we will focus on the case where the random variable is a control for the cost operator of the QAOA; our results likely can be extended to use the random variable as a control on the mixing operator. Reserve $n_x$ qubits for the first-stage variables, $n_y$ qubits for the second-stage variables, and $n_\xi$ qubits for the random variable. Unless otherwise stated, these registers are written in the order $\ket{x}\ket{y}\ket{\xi}$. Define $n=n_x + n_y + n_\xi$; we leave $x$ as a part of the analysis, but in many problems (especially if the second-stage cost function $q$  is linear), it could be removed from the quantum circuit and its contributions computed classically. 

The objective of this section is to design a process that coherently prepares a wavefunction that, for a candidate $x$ as a fixed bitstring, stores a superposition of each optimal second-stage decision $y_\omega^*$ with its corresponding scenario $\xi_\omega$. Call this the \emph{per-scenario optimized} wavefunction, and write it as
\begin{equation}\label{eq:psistar}
    \ket{\psi_x^*} = \ket{x}\sum_{\omega \in \Omega} \sqrt{p(\omega)}\ket{y_\omega^*}\ket{\xi_\omega}.
\end{equation}
Additionally, we want to define a Hamiltonian $H_Q$ such that the expectation value of the per-scenario optimized wavefunction on this Hamiltonian is the expected value function: $\bra{\psi_x^*}H_Q\ket{\psi_x^*} = \phi(x)$. In Sec.~\ref{sec:qae}, we evaluate this expectation value with Quantum Amplitude Estimation.

First, we construct the cost Hamiltonian $H_Q$, which will serve the role of the cost Hamiltonian $H_C$ described in Sec.~\ref{subsec:qaoa_overview} (the letter Q is meant to refer, notationally, back to the original 2nd stage objective in Eq. \ref{eq:QQQ}).  This operator will act on all $n$ qubits. Define $q_\omega(x,y) = q(x,y,\xi_\omega)$ for shorthand. Define the cost Hamiltonian $H_{q_\omega}$ representing the second-stage cost function $q_\omega$ for a specific  scenario $\xi_\omega$ via
\begin{equation}\label{eq:scenario_q}
    H_{q_\omega}\ket{x}\ket{y} = q(x,y,\xi_\omega)\ket{x}\ket{y}
\end{equation}
and acting on the $n_x + n_y$ qubit register for the $x$ and $y$ decision variables. 

We then define the Hamiltonian $H_Q$ as
\begin{equation}\label{eq:HQ}
    H_Q = \sum_{\omega \in \Omega} H_{q_\omega}\otimes \ket{\xi_\omega}\bra{\xi_\omega}.
\end{equation}
Because we combine the scenario-specific Hamiltonian $H_{q_\omega}$ with a projector onto that scenario, each $H_{q_\omega}$ will be applied to the decision register if and only if the distribution register is in that same scenario. For example, consider applying just a single term in this Hamiltonian, $H_{q_j}\otimes \ket{\xi_j}\bra{\xi_j}$, to two separate scenarios $\xi_j$  and $\xi_k,$ where $ j \neq k$ . It is straightforward to see that $\left( H_{q_j}\otimes\ket{\xi_j}\bra{\xi_j} \right) \ket{x}\ket{y}\ket{\xi_j} = q(x,y,\xi_j)\ket{x}\ket{y}\ket{\xi_j}$ and $\left( H_{q_j}\otimes\ket{\xi_j}\bra{\xi_j} \right) \ket{x}\ket{y}\ket{\xi_k} = 0$. To simplify analysis and make its function clear, the definition of this Hamiltonian is oracular. In practice, it can be implemented efficiently with local operators; this is because the random variable usually interacts with the optimization through objects like constraints, which, as discussed in Sec.~\ref{subsec:qaoa_overview}, almost always can be encoded with local $Z$ rotations. This is why we say the distribution register is the control for a Hamiltonian acting on the optimization register. We give one implementation as a part of our example in Sec.~\ref{sec:example}.

If we apply $H_Q$ to an arbitrary wavefunction, where the $\alpha$ are probability amplitudes normalized as $\sum_{x,y,\omega}|\alpha_{x,y,\omega}|^2 = 1$, we can see that this Hamiltonian distributes:
\begin{equation}\label{eq:distribute_HQ}
    H_Q \sum_x\sum_y\sum_{\omega\in\Omega} \alpha_{x,y,\omega}\ket{x}\ket{y}\ket{\xi_\omega} =\sum_x\sum_y\sum_{\omega\in\Omega} q(x,y,\xi_\omega)\alpha_{x,y,\omega}\ket{x}\ket{y}\ket{\xi_\omega}.
\end{equation}
Therefore, suppose we have a candidate $x$ and the optimal second-stage decision $y^*_\omega$ for each scenario $\xi_\omega$; in other words, the wavefunction $\ket{\psi_x^*}$ from Eq.~\ref{eq:psistar}, above. We can see that its expectation value on $H_Q$ is
\begin{equation}\label{eq:exp_value}
\begin{split}
    \bra{\psi_x^*} H_Q \ket{\psi_x^*} &= \bra{\psi_x^*}\bigg[ \sum_{\omega\in\Omega} H_{q_\omega}\otimes\ket{\xi_\omega}\bra{\xi_\omega}\bigg] \ket{x}\sum_{\omega\in\Omega} \sqrt{p(\omega)}\ket{y_\omega^*}\ket{\xi_\omega}\\
    &= \bra{x}\sum_{\omega\in\Omega} \sqrt{p(\omega)}\bra{y_\omega^*}\bra{\xi_\omega}\bigg[\ket{x}\sum_{\omega \in \Omega} q(x,y,\xi_\omega) \sqrt{p(\omega)} \ket{y^*_\omega}\ket{\xi_\omega}\bigg]\\
    &= \sum_{\omega\in\Omega} p(\omega) q(x,y_\omega^*,\xi_\omega)\\
    &= \phi(x),
\end{split}
\end{equation}
which is the expected value function. It is important to note that $H_Q$ does not depend on how $\xi$ distributes probabilities, since $H_Q$ only relies on $\xi$ for control logic; instead, the probability mass of a given outcome $\omega$ is stored as the square of the amplitude on $\ket{\xi_\omega}$ in the wavefunction $\ket{\xi}$.  We apply the Hamiltonian with matrix exponentiation, $U_Q(\gamma) = e^{-i\gamma H_Q}$.

\begin{figure}[t]
    \centering
    \includegraphics[trim={.9cm .1cm 0 0},clip,width=0.70\linewidth]{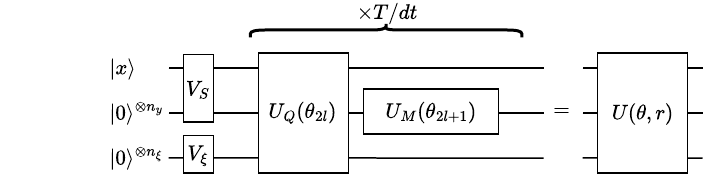}
    \caption{
    \label{fig:qaoa} The general form of DQA with the QAOA, utilizing auxilliary registers for the first-stage decision and the discrete random variable. The cost Hamiltonian $H_Q$ includes information from all three registers, and the mixing Hamiltonian $H_{M_y}$ will modify the $y$ register to find the smallest energy state available, which will encode the expected value function.}
\end{figure}

Now we define the state preparation stage of DQA. Recall that we are given a first-stage decision $x$ as an input. Suppose we have an operator $V_{\xi}$, which prepares the random variable $\xi$ as a wavefunction:
\begin{equation}
V_\xi \ket{0}^{\otimes n_\xi} = \sum_{\omega \in \Omega} \sqrt{p(\omega)} \ket{\xi_\omega} = \ket{\xi}.
\end{equation}
This is possible in general and efficient for specific distributions~\cite{grover2002creating, Huang_2022, dasgupta2022loading, Plesch_2011, Zoufal_2019}. Suppose we are constrained to a subspace $S$, where the constraint involves both first and second stage variables. We prepare the state with
\begin{equation}
    V_S\ket{x}\ket{0}^{\otimes n_y} = \ket{x}\left[ \frac{1}{\sqrt{|S|}}\sum_{y\in S}\ket{y}\right].
\end{equation}
For example, consider a constraint where the $x$ and the $y$ registers need to have total Hamming weight $k$; because $x$ is just a fixed bitstring, we put the $y$ register into a uniform superposition of all states that satisfy $\text{Ham}(x)+\text{Ham}(y) = k$. 

Including the wavefunction representing the random variable, we prepare the initial wavefunction for DQA with 
\begin{equation}\label{eq:initialize}
\begin{split}
    \Big(V_S\otimes & V_\xi \Big)\ket{x}\ket{0}^{\otimes n_y}\ket{0}^{\otimes n_\xi} \\
    & =\ket{x}\left( \frac{1}{\sqrt{|S|}}\sum_{y\in S}\ket{y}\right)\left(\sum_{\omega \in \Omega} \sqrt{p(\omega)} \ket{\xi_\omega} \right).
\end{split}
\end{equation}
Define the mixing operator to \textit{only} operate on the $y$ register, and leave the other registers constant, 
\begin{equation}
    U_{M_y}(\beta) = \mathbb{I}_{2^{n_x}} \otimes U_{M}(\beta) \otimes \mathbb{I}_{2^{n_\xi}},
\end{equation}
where $U_M$ is the mixer described in Sec.~\ref{subsec:qaoa_overview}. Combined with the initialization in Eq.~\ref{eq:initialize} and cost Hamiltonian in Eq.~\ref{eq:HQ}, we can create our modified DQA:
\begin{equation}\label{eq:saqaoa}
    U(\theta, r) = \left[ \prod_{l=1}^r U_{M_y}(\theta_{2l+1}) U_Q(\theta_{2l})\right] (V_S\otimes V_\xi).
\end{equation}
See Fig.~\ref{fig:qaoa} for a circuit diagram of the ansatz. We write the wavefunction prepared by this algorithm as 
\begin{equation}\label{eq:trail_qaoa}
\begin{split}
    \ket{\tilde{\psi}_x^*}& = U(\theta, r) \ket{x}\ket{0}^{\otimes n_y}\ket{0}^{\otimes n_\xi}  \\
    &=\ket{x}\left[\sum_{\omega\in\Omega}\sqrt{p(\omega)}\left(\alpha_{\omega,y^*_\omega}\ket{y_\omega^*} + \sum_{y\neq y_\omega^*} \alpha_{\omega,y}\ket{y} \right)\ket{\xi_\omega} \right],
\end{split}
\end{equation}
where the $y$ register is the one being optimized, and again $\alpha$ are normalized probability amplitudes. This state is an approximation of the wavefunction $\ket{\psi_x^*}$; as the annealing time increases, the state becomes a better approximation, and each $\alpha_{\omega, y_\omega^*}\rightarrow 1$. The parameter $\theta$ is a $2r$ dimensional vector of angles where $\theta_{2l} = a(l*dt)$ and $\theta_{2l+1} = b(l*dt)$ are the interpolating functions introduced in Sec.~\ref{subsec:qaoa_overview}. Write the expectation value of the state prepared in Eq.~\ref{eq:trail_qaoa} as $\bra{\tilde{\psi}_x^*} H_Q \ket{\tilde{\psi}_x^*} \equiv \langle H_Q \rangle_\xi$. By Eq.~\ref{eq:exp_value} and the variational principle, we know that $\langle H_Q\rangle_\xi \geq \phi(x)$. In terms of a ``residual temperature'' $\delta$ we can say that
\begin{equation}\label{eq:variational_error}
    \langle H_Q \rangle_\xi = \phi(x) + \delta,\; \delta\geq 0.
\end{equation}
By the adiabatic theorem, we expect that $\lim_{T\rightarrow\infty} \delta = 0$, and in the adiabatic limit $\delta$ generally decreases as $\delta\sim1/T^2$~\cite{Suzuki_2005}. See Appendix~\ref{appendix:residual_temp} for a more detailed expression of Eq.~\ref{eq:variational_error}. In Sec.~\ref{subsubsec:adiabatic_stable_state}, we show that, assuming $\ket{\xi}$ can be prepared efficiently, the procedure outlined here can produce a state $\ket{\tilde{\psi}_x^*} \approx \ket{\psi_x^*}$ in time independent of $n_{\xi}$ and, by extension, the size $|\Omega|$ of the sample set $\Omega$.

\subsection{Adiabatic evolution of a system coupled to a stable register}\label{subsubsec:adiabatic_stable_state}
In this section, we will discuss the process outlined above in Sec.~\ref{subsec:pdfqaoa} in the language of quantum annealing (QA)~\cite{farhi2000quantum, Albash_2018}. We do this because QAOA and QA have related sources of computational hardness~\cite{farhi2014quantum, kapit2024approximability} (one inspired the other), and thus we expect the arguments to generalize to our case, where we use a QAOA to implement a digitized (or discretized time evolution) QA. We consider the case where only some of the system is included in the driving operator.  This allows us to separate consideration of the second stage optimization variable $y$ from the random variable $\xi$. If we have a computational register for the second-stage variables $y$ and a distribution register to store the random variable $\xi$ as a superposition of its possible scenarios, we argue that (1) QA can prepare the per-scenario optimized wavefunction, which is a superposition of second-stage decision/scenario pairs, and (2) QA can do this with time only proportional to the size of the computational register. We make this argument by considering specific cases of the Hamiltonians described in Sec.~\ref{subsec:pdfqaoa} that generalize easily. We assume the probability mass is independently identically distributed (i.i.d.) across the sample set; i.e. is a uniform superposition $\ket{\xi} = \ket{+}^{\otimes n_\xi}$ of $2^{n_\xi}$ scenarios. This assumption is also easily generalized. We disregard the $x$ component for simplicity (set $n_x = 0$), as its contribution can be considered as a special case of $\xi$, which will be apparent later in this section.

Assume we have a set of $n$ qubits, partitioned into sets $A$ and $B$ such that $A\cup B = \{1,2,\dots,n\}$ and $A\cap B = \emptyset$. We will refer to $A$ as the ``computational'' register for the variables $y$ with $|A| = n_y$, and $B$ as the ``distribution'' register for the random variable $\xi$ with $|B|=n_\xi$. Define phase Hamiltonian $H_Q = \sum_{j<k} J_{jk} Z_jZ_k + \sum_j h_j Z_j$ and driving Hamiltonian $H_M = \sum_{j\in A} X_j$, where $J$ includes interaction strengths across the registers $A$ and $B$, and $H_M$ only acts on the computational register $A$. All qubits are initialized in an even superposition, $\ket{\psi(s=0)} = \ket{+}^{\otimes n}$, which is the ground state of $H_M$ combined with the i.i.d. random variable in the distribution register. Each possible measurement outcome of the distribution register is a single scenario $\xi_\omega$ with $p(\omega)=1/2^{n_\xi} $. The system is evolved by the Hamiltonian 
\begin{align}
    H(s) = (1-s)H_M + sH_Q
\end{align}
as $s$ goes from $0$ to $1$ over time $T$. If $T$ is sufficiently long, the system will remain in the ground state of the instantaneous Hamiltonian $H(s)$ until $s=1$, where $H(s=1) = H_Q$. The operator $U(\theta, r)$ in Eq.~\ref{eq:saqaoa} is mimicking this process on a digital quantum computer, with $r$ chosen to follow the annealing time $T$ and $\theta$ to track $s$ over its evolution (in this case, $a$ and $b$ are chosen to be linear interpolations between $0$ and $1$). 

First, let us show how the distribution register influences the computational register. These ideas follow from other works like~\cite{Hen_2014}. By examining how the term $Z_0Z_1$ functions in a simple 2-qubit system, when qubit $0\in A$ and qubit $1\in B$, we can deduce the expected behavior of this adiabatic evolution. Notice that 
\begin{equation}\label{eq:decompose_zz}
    Z_0Z_1 = Z_0\otimes \ket{0}\bra{0}_1 - Z_0\otimes \ket{1}\bra{1}_1 
\end{equation}
and let the problem and mixing Hamiltonians be $H_Q=Z_0Z_1$ and $H_M = X_0$ ($H_M$ only acts on the $0$th qubit because it is the computational register). First, consider starting in the state $\ket{+0}$ and evolving with $H(s) = (1-s)X_0 + sZ_0Z_1$ over sufficiently long $T$. By the definition in Eq.~\ref{eq:decompose_zz}, this is equivalent to starting in the state $\ket{+}$ and evolving under $H(s) = (1-s)X_0 + sZ_0$, as both result in a $\ket{1}$ in the 0th qubit. Similarly, starting in the state $\ket{+1}$ and evolving with $H(s) = (1-s)X_0 + sZ_0Z_1$ is equivalent to starting in the state $\ket{+}$ and evolving with $H(s) = (1-s)X_0 - sZ_0$, because both result in a $\ket{0}$ in the 0th qubit. Therefore, if we start in the state $\ket{++}$ and evolve under $H(s) = (1-s)H_M + sH_Q$, after a sufficiently long time we expect the final state of the adiabatic evolution to be $1/\sqrt{2} \left( \ket{10} +\ket{01}\right)$. Observe that each $Z$ basis vector of qubit $1$ (the distribution register) dictates a different ground state in qubit $0$ (the computational register).\footnote{While the two ground states of $ZZ$ are degenerate, this not necessarily the case in arbitrary problem Hamiltonians, and also is not the reason we get this result here.} Generalizing this observation to the full system, and noting that the Hamiltonian $H_Q$ is a complex spin-glass, we expect that each scenario (bitstring in the distribution register) will lead to its own ground state in the computational register; in other words, we get a superposition of each scenario in the random variable coupled with the best decision in the computational register for that scenario:
\begin{equation}
    \ket{\psi(s=1)} = \frac{1}{\sqrt{2^{n_\xi}}} \sum_{\omega=0}^{2^{n_\xi}-1} \ket{y_\omega^*}\ket{\xi_\omega}
\end{equation}
where, as stated before, $\xi$ is i.i.d. over all $2^{n_\xi}$ scenarios. This is why we expect an adiabatic evolution of this type to prepare the per-scenario optimized wavefunction discussed in Sec.~\ref{subsec:pdfqaoa}.

Now that we know the final state of the adiabatic evolution, we want to find what the annealing time $T$ should be (asymptotically) to ensure adiabaticity (which ensures this optimization scheme performs well). For QA to remain in the ground state of $H(s)$, we must evolve $H(s)$ slower than the size of the minimum energy gap, $\Delta_{\min}$, between the ground state and first excited state of  $H(s)$. This leads to an annealing time of $T \propto 1/\Delta_{\min}^2$. Therefore, if we can upper-bound the minimum gap of $H(s)$, we can estimate the required time to prepare the per-scenario optimized wavefunction.

We examine this gap in the context of unstructured search, which is an extreme version of optimization problems. Some realistic problems have a slightly larger, but still exponential, minimum gap. Following the argument used in~\cite{ farhi1996analog,PhysRevA.68.062312} the minimum gap of an unstructured search adiabatic evolution is proportional to the overlap between the initial and final states: $\Delta_{\min} \sim |\bra{\psi(s=1)}\ket{\psi(s=0)}|$.  Remember that $|A| = n_y$, $|B|=n_\xi$, and $n=n_y+n_\xi$. If we start in an even superposition
\begin{equation}\label{eq:initial_adiabatic_wvfn}
\begin{split}
    \ket{\psi(s=0)} &= \ket{+}^{\otimes n} \\
     &= \left(\frac{1}{\sqrt{2^{n_y}}} \sum_{y=0}^{2^{n_y}-1} \ket{y}\right)\otimes \left(\frac{1}{\sqrt{2^{n_\xi}}} \sum_{\omega=0}^{2^{n_\xi}-1} \ket{\xi_\omega}\right) \\
     &= \left(\frac{1}{\sqrt{2^{n}}} \sum_{y=0}^{2^{n_y}-1} \sum_{\omega=0}^{2^{n_\xi}-1} \ket{y}\ket{\xi_\omega}\right),
\end{split}
\end{equation}
we can compute the overlap between the two wavefunctions as
\begin{equation}
\begin{split}
    \bra{\psi(s=1)}\ket{\psi(s=0)} &=
    \left[\frac{1}{\sqrt{2^{n_\xi}}}\sum_{\omega=0}^{2^{n_\xi}-1}\bra{y_\omega^*}\bra{\xi_\omega} \right] \left[\frac{1}{\sqrt{2^{n}}} \sum_{y=0}^{2^{n_y}-1} \sum_{\omega=0}^{2^{n_\xi}-1} \ket{y}\ket{\xi_\omega}\right]\\
    &=\frac{1}{\sqrt{2^{n_\xi}}\sqrt{2^n}} \sum_{\omega=0}^{2^{n_\xi}-1}\bra{y_\omega^*}\bra{\xi_\omega}  \left(\ket{y_\omega^*}\ket{\xi_\omega}\right)\\
    &= \frac{1}{\sqrt{2^{n_\xi}}\sqrt{2^n}} \sum_{\omega=0}^{2^{n_\xi}-1} 1
    ,
\end{split}
\end{equation}
where we get the last line above because $\bra{y_\omega^*}\bra{\xi_\omega}(\ket{y}\ket{\xi_\omega}) = \delta(y,y_\omega^*)$, and here $\delta$ is the Kronecker delta. Simplifying further, 
\begin{equation}
\begin{split}
    \bra{\psi(s=1)}\ket{\psi(s=0)} &= \frac{2^{n_\xi}}{\sqrt{2^{n_\xi}}\sqrt{2^n}} \\
    &= \frac{\sqrt{2^{n_\xi}}}{\sqrt{2^n}} \\
    &= 2^{n_\xi/2}  2^{-n/2} \\
    &= 2^{-n_y/2} \\
    &= \frac{1}{\sqrt{2^{n_y}}}.
\end{split}
\end{equation}
The overlap between the two states is therefore only proportional to the size $n_y$ of the computational register, and by~\cite{farhi1996analog} an annealing time $T \sim \sqrt{2^{n_y}}$ is sufficient to remain in the instantaneous ground state of $H(s)$ for a search problem. Thus, we say that when QA uses a distribution register to inform optimization with respect to a random variable, we can safely expect the size of this register, and by extension the number of scenarios encoded in its wavefunction (the size of the sample set), to have no direct contribution to our annealing time. As noted above, we can assume this scaling extends to DQA, the digital quantum computing ``analog'' of QA. This supports the claims made in Sec.~\ref{subsec:pdfqaoa}.

\section{Calculating the expectation value with Quantum Amplitude Estimation}\label{sec:qae}
We now discuss how Quantum Amplitude Estimation (QAE) can produce an estimate of the expectation value. Picking up where the last section left off, assume we can prepare a wavefunction $\ket{\tilde{\psi}_x^*} = U(\theta,r)\ket{x}\ket{0}^{\otimes n_y+n_\xi}$. This wavefunction has expectation value $\langle H_Q \rangle_\xi $ on the Hamiltonian $H_Q$, which is related to the true expected value function as $\langle H_Q\rangle_\xi = \phi(x) + \delta$, and we know $\lim_{T\rightarrow \infty}\delta = 0$. We use QAE to produce an estimate of the expectation value $\langle H_Q\rangle_\xi$ in time polynomially faster than Monte Carlo sampling. We first describe canonical QAE with the assumption that the wavefunction has converged to the per-sample optimized wavefunction, $\ket{\psi_x^*}$, for a candidate $x$. Then, we discuss the overall error between our estimate of the expected value function and the true expected value function in Sec.~\ref{subsec:error}. 

We will describe canonical QAE, as presented in~\cite{Brassard_2002}. Many variants of this algorithm exist, with different scaling, convergence, and noise-tolerance~\cite{Giurgica_Tiron_2022, Suzuki_2020, PhysRevApplied.15.034027}; we focus on the canonical algorithm to simplify the analysis. QAE works by changing a sampling problem into a period estimation problem; we describe how QAE does this, and then discuss implementation details.

Suppose we have the operator $\mathcal{A}$, defined in detail later, acting on $n$ system qubits and a single ancilla qubit as
\begin{equation}\label{eq:A}
    \mathcal{A}\ket{0}^{\otimes n}\ket{0} = \sqrt{1-a}\ket{\psi_0}\ket{0} + \sqrt{a}\ket{\psi_1}\ket{1},
\end{equation}
where $\ket{\psi_0}$ and $\ket{\psi_1}$ are arbitrary normalized wavefunctions, and we want to obtain an estimate of the value $a$; i.e. the probability amplitude on the $\ket{1}$ state of the ancilla qubit. In classical Monte Carlo sampling, we simply measure $M$ samples of the ancilla qubit and count the number of $\ket{1}$'s observed, and the error of this estimate converges in $O(1/\sqrt{M})$. To utilize QAE, we make an operator such that $a$ is given by the phase of its eigenvalue. This is the Grover operator 
\begin{equation}\label{eq:Q}
    \mathcal{Q} = \mathcal{A}S_0\mathcal{A}^\dagger S_{\psi_0}
\end{equation}
where $S_0 = \mathbb{I} -2 \ket{0}_{n+1}\bra{0}_{n+1}$ and $S_{\psi_0} = \mathbb{I} -2 \ket{\psi_0}\ket{0}\bra{\psi_0}\bra{0}$. Quantum Phase Estimation (QPE)~\cite{kitaev1995quantum} is then used to estimate the eigenvalues of $\mathcal{Q}$; it is known that the eigenvalues of $\mathcal{Q}$ correspond to an estimate of $a$ (refer to~\cite{Brassard_2002}). This estimate has error $O(1/M)$, where now $M-1$ is the number of times the operator $\mathcal{Q}$ has to be repeated for the QPE algorithm. 

\begin{figure*}[t]
    \centering
    \includegraphics[trim={.9cm 0cm 0 0},clip, scale=.9]{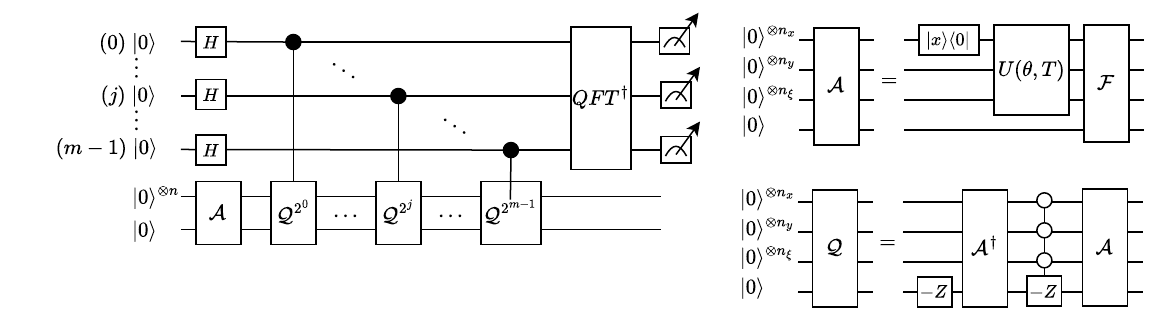}
    \caption{Circuit schematic of QAE. The $\mathcal{A}$ and $\mathcal{Q}$ subroutines are shown on the right. The operator $\ket{x}\bra{0}$ initializes the first-stage register in the classical state $x$. Empty circles on a control qubit indicate control by the $\ket{0}$ state.}
    \label{fig:qae}
\end{figure*}

The unitary operator $\mathcal{A}$ is $\mathcal{A} = \mathcal{F}\left(U(\theta,T)\otimes \mathbb{I}\right)$, where $U(\theta,T)$ is the DQA described in Sec.~\ref{sec:aqaoa}, and $\mathcal{F}$ is an oracle operator, which we now describe. This operator takes the $n$ qubits for $x$, $y$, and $\xi$, and an additional ancilla qubit, and computes the normalized value of the second-stage objective function into the probability amplitude of the ancilla qubit. Recall from Sec.~\ref{sec:intro} that we have a lower and upper bound on the function $q$, such that $q_l \leq q(\bullet,\bullet, \bullet) \leq q_u$. This allows us to define $\bar{q}: \{0,1\}^{n_x+n_y+n_\xi}\rightarrow [0,1]$, which means we can interpret the square root of the value of $q(x,y,\xi_\omega)$ as a probability amplitude as seen in Eq.~\ref{eq:A}, and its value as the probability of measuring the ancilla qubit in the $\ket{1}$ state. More explicitly, we set 
\begin{equation}\label{eq:qbar}
    \bar{q}(x,y,\xi_\omega) = \frac{q(x,y,\xi_\omega) - q_l}{q_u - q_l}.
\end{equation}
The operation of the oracle operator is then defined as
\begin{equation}\label{eq:oracle}
\mathcal{F}\ket{x}\ket{y}\ket{\xi_\omega}\ket{0} = \ket{x}\ket{y}\ket{\xi_\omega}\times\left(\sqrt{1-\bar{q}(x,y,\xi_\omega)} \ket{0} + \sqrt{\bar{q}(x,y,\xi_\omega)}\ket{1}\right),
\end{equation}
which means if we apply it to $\ket{\psi_x^*}$, we get
\begin{equation}
\begin{split}
    \mathcal{F}\ket{\psi_x^*}\ket{0} &= \sum_{\omega\in\Omega} \sqrt{p(\omega)} \ket{x}\ket{y_\omega^*}\ket{\xi_\omega}\left( \sqrt{1-\bar{q}(x,y_\omega^*,\xi_\omega)} \ket{0} + \sqrt{\bar{q}(x,y_\omega^*,\xi_\omega)}\ket{1} \right)\\
    &= \sqrt{1-\bar{\phi}(x)}\ket{\psi_0} \ket{0} + \sqrt{\bar{\phi}(x)}\ket{\psi_1}\ket{1} .
\end{split}
\end{equation}
And the probability of measuring a $\ket{1}$ in the ancilla qubit is
\begin{equation}\label{eq:prob1}
\begin{split}
    \text{Pr}\left[\ket{1}\right] = \sum_{\omega\in\Omega}p(\omega)\bar{q}(x,y_\omega^*,\xi_\omega) = \bar{\phi}(x),
\end{split}
\end{equation}
where $\bar{\phi}(x) = \left(\phi(x) - q_l\right)/\left(q_u - q_l\right)$. Therefore, when the DQA has produced $\ket{\psi_x^*}$, we know that $a$ in Eq.~\ref{eq:A} is $\bar{\phi}(x)$; in general, we have $\text{Pr}\left[\ket{1}\right] = a = \left( \langle H_Q\rangle_\xi - q_l\right)/\left(q_u-q_l \right)$ (see Appendix~\ref{appendix:oracle_res_temp}).

Implementing $\mathcal{F}$ usually involves some complex sequence of controlled-$Y$ rotations, based on the observation that  $RY(2\arcsin(\sqrt{\bar{q}}))\ket{0} = \sqrt{1-\bar{q}}\ket{0} + \sqrt{\bar{q}}\ket{1}$. However, because $q$ will in general take a different value for different inputs, and $RY(2\arcsin(\sqrt{\bar{q}_1}))RY(2\arcsin(\sqrt{\bar{q}_2})) \neq \sqrt{1-\bar{q}_1-\bar{q}_2}\ket{0} + \sqrt{\bar{q}_1+\bar{q}_2}\ket{1}$, constructing $\mathcal{F}$ requires some consideration. There are many different ways of implementing $\mathcal{F}$, which usually involve a tradeoff between accuracy, time complexity, and ancilla qubits. Near-term algorithms utilize the small-angle approximation~\cite{Stamatopoulos_2020, Woerner2019}, while fault-tolerant algorithms will likely compute $\arcsin{\sqrt{\bar{q}}}$ into an ancilla register first with quantum arithmetic~\cite{bhaskar2015quantum, PhysRevA.54.147} and then compute the probability amplitude to the ancilla qubit, as described in~\cite{Chakrabarti_2021}. Examining these different oracles is beyond the scope of this work, and we refer the reader to the above sources for more information. We implement two different oracles as a part of the example in Sec.~\ref{sec:example}.

Putting these components together, the QAE routine works as follows. A circuit diagram is shown in Fig.~\ref{fig:qae}. We refer to the $m$ control qubits used for phase estimation as ``estimate'' qubits, and the other $n+1$ qubits that house the optimization problem and ancilla qubit as the ``system'' qubits. First, we reserve $m$ estimate qubits and initialize these qubits into an even superposition with $H^{\otimes m}$. At the same time, we apply the operator $\mathcal{A}$ to the system qubits, which involves running the DQA routine $U(\theta,T)$ from Sec.~\ref{sec:aqaoa} and applying the oracle $\mathcal{F}$  to place amplitude $\sqrt{a}$ on the $\ket{1}$ state of the ancilla qubit. Now for each qubit $j$ in the $m$ sample qubits, we do a controlled operation $C\mathcal{Q}^{2^j}$, which applies the operator $\mathcal{Q}$ on the system qubits a total of $2^j$ times, controlled by estimate qubit $j$. Notice that we zero-index the sample qubits, so the first qubit is $0$, with $2^0=1$ applications of $\mathcal{Q}$. Finally, we perform an inverse Quantum Fourier Transform $QFT^\dagger$~\cite{coppersmith2002approximate} on the estimate qubits, and measure the resulting bitstring $b$ in the $m$ sample qubits.

Taking the result $b$ as its integer value, we compute 
\begin{equation}
    \tilde{a} = \sin^2\left(\frac{b\pi}{2^m}\right),
\end{equation}
which is our normalized estimate of the expected value function. We compute our estimate as $\tilde{\phi}(x) = \tilde{a}*(q_u-q_l) +q_l$. Note that the length $m$ of the bitstring $b$ implies a certain discretization of the possible values of its equivalent integer, which in turn discretizes our estimate $\tilde{a}$.

This estimate has the error~\cite{Brassard_2002,Montanaro_2015}
\begin{equation}\label{eq:qae_error}
    \text{Pr}\left[|\tilde{a} - a| \leq \frac{\pi}{M} + \frac{\pi^2}{M^2} \right] \geq \frac{8}{\pi^2} \approx{0.81},
\end{equation}
which is where we derive the error convergence rate of $O(1/M)$, a polynomial advantage over the classical Monte Carlo convergence of $O(1/\sqrt{M})$~\cite{glasserman2000}. The $8/\pi^2$ is a part of the canonical error bound of QAE~\cite{Brassard_2002}; we overcome this by repeating the experiment as many times as necessary (i.e., to reduce the probability of failure rapidly from $\approx .19$ to $.19^2, 0.19^3 \rightarrow 0$), and in general this failure probability can be made arbitrarily small with a multiplicative factor of $O(1/\log(\gamma))$ to get success probability $1-\gamma$~\cite{JERRUM1986169}.

Remember that $a=(\langle H_Q\rangle_\xi - q_l)/ (q_u - q_l)$. This is because, in general, the state prepared by DQA will not have converged perfectly to the ground state; i.e. $\langle H_Q \rangle_\xi = \phi(x) + \delta$. QAE estimates $\langle H_Q\rangle_\xi$ with the residual temperature and not the exact, underlying expected value function. Additionally, we would like to remark that $\langle H_Q\rangle_\xi$ can still be normalized to $a\in[0,1]$ with the same $q_l,q_u$ by Eqs.~\ref{eq:distribute_HQ},\ref{eq:trail_qaoa} (again, see Appendix~\ref{appendix:oracle_res_temp}).

\subsection{Time to estimate with given accuracy and precision}\label{subsec:error}
Since QAE requires $2^{m+1}-1$ repetitions of the operator $\mathcal{A}$ to compute an estimate onto the $m$ estimate qubits, it is important to examine the number of estimate qubits required to achieve a sample with bounded
error, especially since $\mathcal{A}$ itself may have circuit depth exponential in $n_y$ in the worst case. This section looks at the number of times we have to repeat $\mathcal{A}$ in order to get an estimate of the expected value function with additive error $|(\tilde{\phi}(x) - \langle H_Q \rangle_\xi )/(q_u - q_l)| \leq \epsilon$ with high probability, as well as how close we can expect the output of the algorithm to be to the true expected value function. 

Following the argument put forward by Montanaro in~\cite{Montanaro_2015}, we can use QAE to get an estimate within additive error $\epsilon$ in $O(1/\epsilon)$ repetitions of $\mathcal{A}$ if we can place a bound on our second-stage function $Q$. This is because, if the second-stage objective $Q$ is bounded (which it is, as we bound the objective function $q$ that $Q$ minimizes), then the variance is also bounded. Referring to Eq.~\ref{eq:prob1}, the operator $\mathcal{F}$ then turns our random variable (now $q$) into a Bernoulli Random Variable (BRV); the expected value of this variable is what QAE estimates. Since it is a BRV, it has variance $\leq 1/4$ (by Popoviciu's inequality on variances~\cite{Popoviciu}). Then, to get an estimate with additive error less than $\epsilon$, we require $O(1/\epsilon)$ repetitions of $\mathcal{A}$. This only has asymptotic dependence on target additive error $\epsilon$.  Thus, the number of repetitions $M$ of $\mathcal{A}$ needed to perform QAE does not depend on the size of the sample space $\Omega$. 

Our algorithm computes $\tilde{\phi}(x)$. We calculate the full two-stage objective function using this estimate of the expected value function, via $\tilde{o}(x) = f(x) + \tilde{\phi}(x)$. By looking at the error estimate from Eq.~\ref{eq:qae_error}, substituting the expectation value for the true expected value function, $\langle H_Q \rangle_\xi = \phi(x) + \delta$, and including the normalization, we can write the final error of $\tilde{\phi}(x)$ as calculated by our algorithm as
\begin{equation}\label{eq:error_bound}
    \text{Pr}\left[\left|\frac{\tilde{\phi}(x) - \phi(x) - \delta}{q_u - q_l}\right| \leq \frac{\pi}{M} + \frac{\pi^2}{M^2}\right] \geq \frac{8}{\pi^2}.
\end{equation}
The residual temperature in general decreases with $\delta \propto 1/T^2$ when $T$ is in the adiabatic limit~\cite{Suzuki_2005}. This residual temperature also decreases differently for different problems: see~\cite{kapit2024approximability} for a robust discussion of approximation hardness. Notice that the two error sources are additive, where longer DQA time $T$ leads to a better estimate accuracy, and larger QAE depth $M$ leads to better precision. These two circuit depths multiply by each other, to produce an overall runtime $\tilde{O}(TM/dt)$; to reiterate, $T$ only depends on the number of second-stage decision variables, and $M$ only depends (asymptotically) on the target precision. This runtime assumes efficient state preparation for the DQA. Different optimization problems will also introduce different multiplicative overheads to this runtime assessment, as some have more straightforward Hamiltonian encodings than others.

\section{Implementation for binary stochastic unit commitment}\label{sec:example}
We implement our algorithm on an optimization problem inspired by stochastic unit commitment~\cite{HABERG201938, 9281609}. In this problem, the objective is to decide the output of a single traditional generator (say, a gas generator), which must be ``committed" ahead of a future target time, while accounting for uncertainty in renewable energy generation (say, wind turbines, where the uncertainty is in whether the wind will be blowing at the target time).  We simplify the formulation by assuming the amount of gas generation, $x$, is an integer, that the wind turbines can only be on or off, meaning the $y$ variable is a binary vector of length $n_y$, and that wind either blows or it does not, meaning samples of the random variable $\xi$  are binary vectors of length $n_{\xi}$, which in this case is equal to $n_{y}$ (one binary ``wind speed" for each turbine).  In this example, $\Omega = \{\text{wind doesn't blow},\text{wind blows}\}^{n_y}$. We define the random variable at a single turbine as $\xi_{\text{wind doesn't blow}} = 0$ and $\xi_{\text{wind blows}} = 1$. Consider the objective function
\begin{align}\label{eq:uc}
    \min_{\substack{x\in \mathbb{Z}_{\geq 0}}} o(x) = c_xx + \mathbb{E}_\xi[Q(x,\xi)], 
\end{align}
where $x$ is a single positive integer representing the power output from gas generators with linear cost $c_x$. The second stage objective function, defined over $n_y$ wind turbines, is given by 
\begin{align}\label{eq:loss-function}
&Q\left(x,\xi\right) =  \nonumber \\ 
\min_{\substack{y\in \{0,1\}^{n_y}}}\quad  & \sum_{j=0}^{n_y-1} c_j y_j\xi_{\omega,j} - c_r y_j(\xi_{\omega,j}-1)   \nonumber \\
\text{s. t.} \quad  
& 1^T y + x = d.
\end{align}
The binary decision vector $y\in \{0,1\}^{n_y}$ and a single realization of the random vector $\xi_\omega$ are binary vectors with the same dimension. Wind turbine $j$ has linear cost $c_j<c_x$. We keep a hard constraint $1^Ty + x = d$, where $d$ is the power demand. The cost of \textit{not} delivering a unit of power is $c_r$, where $c_r > c_x$ to ``incentivize'' using more gas power when there is not enough wind; we add this value if we rely on a wind turbine $j$ to deliver power for a given scenario and it is unable to, since this coupled with the hard constraint would mean a unit of demand is not satisfied, which is accounted for in the term $-c_ry_j(\xi_j - 1)$. This model is oversimplified from practically useful unit commitment problems, due to constraints on simulating quantum algorithms. To simplify the analysis, we focus on an independently and identically distributed random variable to represent the wind distribution.

\subsection{Circuit formulation}
We form the DQA operator $U(\theta, r)$ as described in Sec.~\ref{sec:aqaoa}. We reserve $n_y$ qubits for the wind turbines, and $n_\xi=n_y$ qubits for the random variable; let the qubits $[0,n_y-1]$ be the wind register and $[n_y,2n_y-1]$ be the random variable register, and the wind at qubit $j+n_\xi$ corresponds to the wind present at the $j$th turbine. The $j$th qubit in the register $n_\xi$ will be in the $\ket{1}$ state if the wind is blowing enough to use the $j$th wind turbine, and $\ket{0}$ otherwise. We do not reserve a register for the first-stage variable; because the second stage functions are all linear it can be separated and accounted for on the classical processor.

We use a hard constraint to satisfy how much power we expect the second stage to deliver: if we set the gas generator at $x$, then we expect the wind turbines to be able to deliver the rest of the power $1^T y = d-x$. This lets us use the constraint preserving mixer:
\begin{equation}
    U_M(\beta) = \prod_{j=0}^{n_y-2}\prod_{k=j}^{n_y-1}\text{SWAP}_{j,k}(\beta),
\end{equation}
which ensures the number of ``1''s in the $y$ register stays equal to $d-x$.  We then use a penalty operator to satisfy the constraint that if we turn on wind turbine $j$ when there is not enough wind, we incur cost $c_r$. $P(\gamma)$ is the phase operator defined as  
\begin{equation}
    P(\gamma) = \begin{pmatrix} 1 & 0 \\ 0 & e^{i\gamma} \end{pmatrix}.
\end{equation}
This lets us define the penalty phase operator, which applied with controlled logic applies a phase proportional to $c_r$ if the turbine qubit $j$ and its wind qubit $j+n_y$ are in the state $\ket{0}_{n_y+j} \ket{1}_j$; i.e. if we try to rely on turbine $j$ to provide wind when none is available. Applied to every turbine qubit, this is
\begin{equation}
    U_P(\gamma) = \prod_{j=0}^{n_y-1} X_{j+n_y} CP_{j+n_y, j}(\gamma c_r) X_{j+n_y}.
\end{equation}
Additionally, we define the cost operator to apply the cost of using turbine $j$ if both it and its wind qubit are on
\begin{equation}
    U_C(\gamma) = \prod_{j=0}^{n_y-1}CP_{j+n_y, j}(\gamma c_j).
\end{equation}
We prepare the initial state to be an even superposition on the distribution register, representing the i.i.d. random variable, $H^{\otimes n_\xi} \ket{0}_{n_\xi} = 1/\sqrt{2^{n_\xi}} \sum_{\omega = 0}^{2^{n_\xi}-1} \ket{\xi_\omega}$ and an even superposition of all states with Hamming weight equivalent to $d-x$ in the $y$ register:  $1/\sqrt{n\choose{d-x}} \sum_{\text{Ham}(y) = d-x} \ket{y}$. This is called the ``Dicke state'', and we prepare it with the unitary operator $V_k$ (where $k$ is the Hamming weight) following the algorithm in~\cite{B_rtschi_2019}. For this specific problem, we can then write the DQA as 
\begin{equation}\label{eq:aqaoa_uc}
\begin{split}
    U_{\text{U.C.}}(T) \ket{0}^{\otimes n} &= \left[\prod_{t=1}^T U_M(1-\frac{t}{T})U_P(\frac{t}{T})U_C(\frac{t}{T})\right]\\
    &\times\left(V_{d-x}\otimes H^{\otimes n_\xi}\right)\ket{0}^{\otimes n}.
\end{split}
\end{equation}
We chose the interpolating functions $a(t)=t/T$ and $b(t)=1-t/T$ with $a(0) = b(T) = 0$ and $a(T)=b(0)=1$. In this case a smaller Trotter step $dt$ is equivalent to a longer annealing time $T$, because the $dt$ appears in both the numerator and denominator and thus cancels. Therefore, in the expression of the ansatz we set $t=l$ and $T=r$.

We also define two different oracle operators, $\mathcal{F}_{\text{exact}}$ and $\mathcal{F}_{\sin}$. The first, $\mathcal{F}_{\text{exact}}$, is prohibitively expensive to implement at scale and just used for demonstration purposes to show QAE convergence. The second, $\mathcal{F}_{\sin}$, is much more practical to implement but may lead to small additional errors in $\tilde{\phi}(x)$ due to the small angle approximation. They have the forms (index $a$ marks the ancilla qubit)
\begin{equation}\label{eq:fexact}
\begin{split}
    \mathcal{F}_{\text{exact}} &= \prod_{y,\omega} \ket{x}\ket{y}\ket{\xi_\omega}\bra{x}\bra{y}\bra{\xi_\omega} \\
    &\hspace{.6in}\otimes RY_a\left(2\arcsin{\sqrt{\bar{q}(x,y,\xi_\omega)}}\right)
\end{split}
\end{equation}
and 
\begin{equation}\label{eq:fsin}
\begin{split}
    \mathcal{F}_{\sin} &= \prod_{j=0}^{n_y-1}CCRY_{j + n_y, j, a}(\pi c_j) X_{j+n_y}\\
    &\hspace{.6in}\times CCRY_{j + n_y, j, a}(\pi c_r)X_{j+n_y}.
\end{split}
\end{equation}
Here, $CCRY_{j, k, a}$ is a doubly controlled rotation onto qubit $a$ with control qubits $j$ and $k$, where the operation $RY_a$ is applied iff both qubits $j$ and $k$ are in the $\ket{1}$ state. Note also, for each run, we have the bounds $q_l=0$ and $q_u=c_r *(d-x)$.

\subsection{Results}
We now show small scaling evidence for the DQA (optimizer convergence without shot noise) and QAE (estimator convergence without residual temperature) components of the algorithm separately and then demonstrate both of them together to optimize the entire two-stage objective function. We select model parameters as follows: gas cost $c_x = 0.4$, wind turbine costs randomly selected in the range $c_j\in [0.01, 0.2]$, recourse cost $c_r = 1.0$, and demand $d=n_y$. All simulations are performed on a classical processor using Qiskit~\cite{Qiskit}.

\begin{figure*}[t!]
    \centering
    \includegraphics[width=1.0\linewidth]{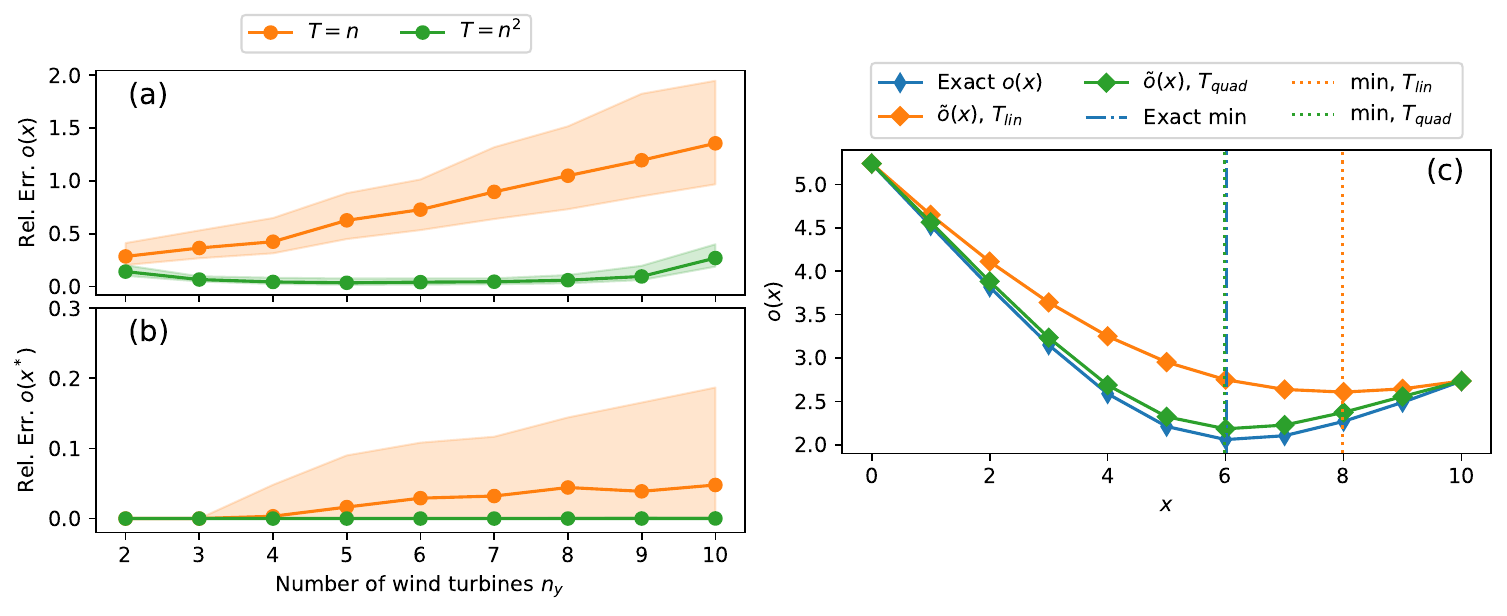}
    \caption{Convergence of the DQA step of our algorithm, using statevector simulation (immune to shot noise, and not requiring QAE). At each size $n_y$, we compute the relative error between the estimated objective function and the true objective function for 30 models. Colored regions are the minimum and maximum of the data. (a) The relative error between the estimated objective function and true objective function, $\sum_{x=0}^{d} |\tilde{o}(x) - o(x)|/o(x)$. (b) The relative error between the objective function at the true minimum and the objective function at the minimum discovered by our algorithm, $|o(\tilde{x}^*) - o(x^*)|/o(x^*)$. (a) and (b) together show that, even when the objective function calculation has high error, an accurate minimum is discovered. (c) A single objective function surface at $n_y=10$, with minima labeled. The $T_{\text{quad}}$ estimator finds the true minimum. The relative error of these curves is $1.65$ (for $T_{\text{lin}}$) and $0.33$ (for $T_{\text{quad}}$).}
    \label{fig:qaoa_results}
\end{figure*}

First, we test the DQA stage of the algorithm. We examine how the error in our ground state estimate, $\delta$, increases in $n_y$ when we choose annealing times $T_{\text{lin}}= n_y$ and $T_{\text{quad}}=n_y^2$, along with how $\delta$ could potentially affect an answer. We choose $T_{\text{lin}}$ because it will show the result of an unconverged optimization, and $T_{\text{quad}}$ to demonstrate a better performing optimization. For a given problem instance, we compute the objective function $\tilde{o}(x)$ over the whole domain $x\in\{0,1,\dots,d\}$. At each value of $x$ in the domain, we use the DQA given by Eq.~\ref{eq:aqaoa_uc}, with either $T_{\text{lin}}$ or $T_{\text{quad}}$, to prepare a trial wavefunction $\ket{\tilde{\psi}_x^*}$. We then compute the expectation value  using that trial wavefunction's state vector, which lets us see convergence towards the true per-scenario optimized wavefunction $\ket{\psi^*_x}$ without any estimator noise, and then compute the objective function $\tilde{\phi}(x)$ at that point with Eq.~\ref{eq:uc}. We then compare this estimated objective function to the true objective function (computed with brute force, i.e., direct enumeration of all possible $y$ values). 

Results for this experiment can be seen in Fig.~\ref{fig:qaoa_results}. At each $n_y$ we repeat for 30 different parameter sets (wind turbine costs). First notice Fig.~\ref{fig:qaoa_results}(a), where we compute relative error between the estimated and true objective functions. Notice that increasing time from $n_y$ to $n_y^2$ greatly reduced the relative error due to the decrease in residual temperature. These times also agree with our hypothesis (in the sense of convergence not depending on the size of distribution) about the annealing time in Sec.~\ref{subsubsec:adiabatic_stable_state}. In Fig.~\ref{fig:qaoa_results}(b) we use the estimated objective function to choose a first-stage decision $\tilde{x}^*$ that we expect to perform well; we compare the objective function at this choice, $o(\tilde{x}^*)$, to the true minimum of the objective function. Notice that this compares the true objective function for two separate $x$ values to check the quality of the solutions. Looking at subplots (a) and (b) together, we see that while the relative error between the objective functions is diverging, the discovered minima are still in the neighborhood of the true minima. The reason for this can be seen in Fig.~\ref{fig:qaoa_results}(c), where we plot the exact objective function against the estimated objective function with $T_{\text{lin}}$ and $T_{\text{quad}}$. While the relative error between the curves is still high, they still follow a very similar shape, and because we study a convex problem here, even minima found with high residual temperature can be in the neighborhood of the true solution. 

\begin{figure*}[t!]
    \centering
    \includegraphics[width=1.\linewidth]{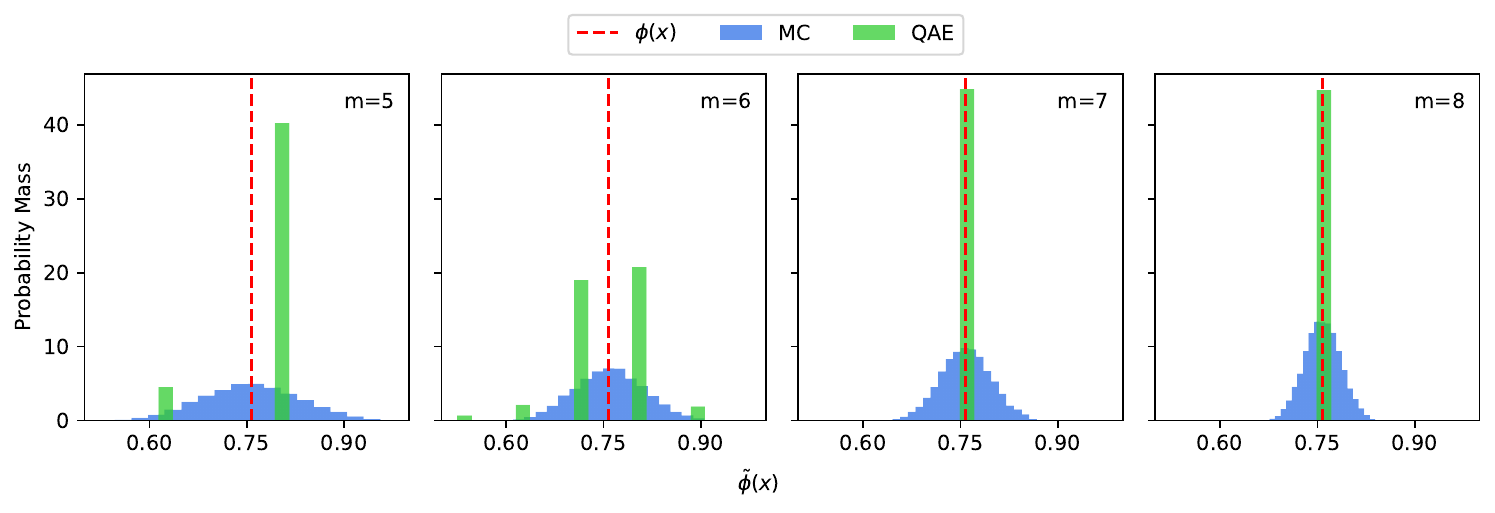}
    \caption{Plotting the density of the estimated expected value functions, for each estimation strategy with $O(2^m)$ samples. For the same wavefunction, we compare QAE and Monte Carlo (MC) sampling to estimate the expected value function (we say MC here, as repeatedly preparing and measuring the wavefunction has MC convergence). With $n_y=3$, we checked $m=5,6,7,8$. We repeated the same estimation task 10,000 times (i.e. formed an estimate with $2^{m+1}$ samples 10,000 times). The horizontal axis is the value of the expected value function, in red, or bins of estimates of it, in blue and green. The vertical axis is the probability mass of the bin at that value (bins have width $\approx 0.02237$). This uses the exact oracle $\mathcal{F}_{\text{exact}}$. As noted in Sec.~\ref{sec:qae}, the estimate can only take certain values dictated by the discretization, which is why the $m=5,6$ panels have zero density overlapping with the true expected value function. We can see here that as $m$ increases, QAE gets both a ``finer grain'' discretization and better likelihood of measuring the true expected value function.}
    \label{fig:qae_results}
\end{figure*}

Second, we examine how canonical QAE performs with the ``brute force'' oracle $\mathcal{F}_{\text{exact}}$ from Eq.~\ref{eq:fexact} using a perfectly converged optimizer, as a function of the number of estimate qubits $m$. This is then compared to a shots-based sampling of the wavefunction, i.e., repeated state preparation and measurement, which has Monte Carlo convergence. These results can be seen in Fig.~\ref{fig:qae_results}. We construct the $\ket{\psi_x^*}$ wavefunction perfectly (to avoid residual temperature), and then repeat 10,000 runs of each sampling technique to estimate $\tilde{\phi}(x)$. For QAE with $m$ sample qubits, we repeat the state preparation algorithm $\mathcal{A}$ (the DQA $U_{\text{U.C.}}(T)$ followed by oracle $\mathcal{F}_{\text{exact}}$) $2^{m+1}+1$ times. Therefore, to properly compare the two sampling techniques, we also estimate the expected value function without QAE, and to make it a fair comparison, we use $2^{m+1}$ measurements. As can be seen in the figure, the QAE method has a higher chance of estimating the true expected value function with fewer samples. This agrees with the theory laid out in Sec.~\ref{sec:qae}. At the smaller values of $m$, an artifact about QAE can be seen: our final result is a bitstring on $m$ qubits, which means the estimated phase is discretized in some way. Because of this discretization, in these plots the green bars are a superposition of the true amplitude; a weighted average of these green bars also would produce the expected value function~\cite{Suzuki_2020}.

\begin{figure*}[ht!]
    \centering
    \includegraphics[width=0.9\linewidth]{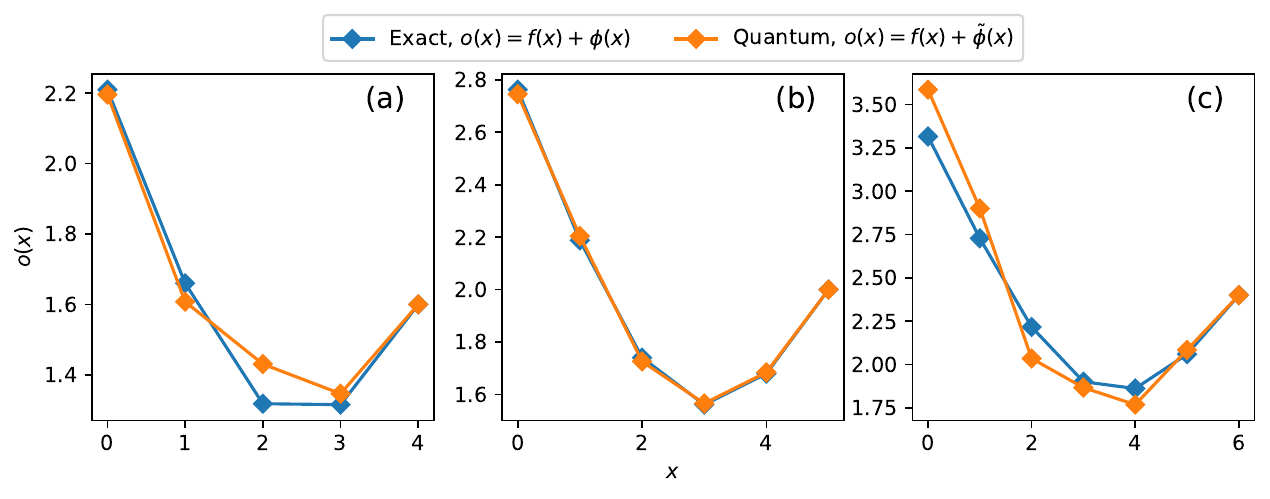}
    \caption{The objective function computed by the entire algorithm, for (a) $n_y=4$, $m=6$, $T=10$; (b) $n_y=5$, $m=6$, $T=15$; (c) $n_y=6$, $m=5$, $T=20$. Each run is with only 1 measurement at the end of the computation (i.e. we don't repeat the algorithm to have a better chance of landing in the $8/\pi^2$ confidence interval). This uses the sin-approximation oracle $\mathcal{F}_{\sin}$.}
    \label{fig:obj_surface}
\end{figure*}

Finally, we show the full algorithm loop by using the quantum algorithm to assist a classical computer in searching for the minimum first-stage decision $x^*$. For three problems, we run the DQA and QAE algorithms for each candidate $x$ (parameters discussed in the caption) and compute the estimated objective function surface. This time, we use the oracle $\mathcal{F}_{\sin}$ from Eq.~\ref{eq:fsin}. The results can be seen in Fig.~\ref{fig:obj_surface}. The objective functions calculated have good agreement with the exact functions and find the true minimum each time.

\section{Discussion and Conclusion}
Our work has not explicitly addressed hardware error, which is a critical issue for modern quantum processors. Canonical QAE is highly vulnerable to noise, because of the length and fragility of QPE. To see utility in a related application (derivative pricing), it is estimated that this algorithm needs fault tolerant quantum computers with $\sim$ 8k qubits~\cite{Chakrabarti_2021}; however, some near term work has had luck implementing the algorithm for toy problems~using more error-resistant implementations~\cite{Woerner2019}. QAOA is seen as a candidate for near-term quantum advantage~\cite{PhysRevX.10.021067, Guerreschi2019}; however, error in the QAOA/DQA component of our algorithm could pose an obstacle for the following reason. Error in the DQA circuit can be seen as additional residual temperature, so if we repeat the DQA multiple times as required by QAE, these errors will start to compound and push the wavefunction further away from the per-scenario optimized wavefunction at each QAE layer due to the exponentially shrinking probability of finding the ground state in the presence of noise in the QAOA circuit~\cite{Pearson_2020}.

The two algorithm components presented here are also intentionally modular;  this future proofs the algorithm for improvements to either QAOA (like the spectral folding algorithm introduced in~\cite{kapit2024approximability}) or QAE (like Maximum Likelihood Estimation~\cite{Suzuki_2020}), as well as allowing for implementations with more noise-resilient component quantum algorithms. We leave examining relative performance improvements for future work. 

We have shown that the Quantum Alternating Operator Ansatz and Quantum Amplitude Estimation algorithms can be used to evaluate the expected cost of an optimization problem solved over a discrete random variable. We show that our estimate can be made in time asymptotically independent from the number of scenarios represented by the discrete random variable. Finally, we implemented this algorithm to solve a simple binary, linear, unit commitment problem. 

More clever tomography techniques could greatly reduce the algorithm runtime. If we can adopt a non-destructive measurement technique that allows state repair, like the one suggested in~\cite{Rall_2023}, we could estimate the expectation value without using QAE. In this case, we would forgo the QAE step of the algorithm, and instead repeatedly apply some partial measurement and state repair to estimate the expected value function. This would mean perform sampling as a post-algorithm routine instead of needing to repeat the operator $\mathcal{Q}$ in Eq.~\ref{eq:Q} $2^m$ times, each of which requires running the QAOA twice. That is, this would take the time complexity from $\tilde{O}(TM/dt)$ to $O(T/dt+M^2)$. This would be useful if the QAOA is prohibitively deep ($T$ is large) for many repetitions in QAE. Additionally, if there is any methodology of performing the operator $\mathcal{Q}$ without running state preparation for the wavefunction, like the random walk technique suggested by~\cite{Harrow_2020}, QAE could still be used, further reducing the runtime to $O(T+M)$. Notably, this is only an additive time scaling away from solving a deterministic optimization problem.

More work needs to be done to examine classical algorithms that can use this procedure as a kernel. We believe that this algorithm is a step towards practically solving a problem class that is both difficult and important, and the proper algorithm to make the first-stage decisions could symbiotically make use of this kernel to solve these programs. Additionally, the way error in the quantum routine causes error in the surface of the objective function, and how this changes answers produced by the classical routine, should be examined formally. This will affect different problems in different ways; in this manuscript, the objective function examined in Sec.~\ref{sec:example} is convex, so error in the quantum algorithm was less likely to move the discovered solution away from the neighborhood around the true solution.

\textbf{Code and data}
Code and data are available at reasonable request.

\textbf{Acknowledgment}
The authors would like to thank Jonathan Maack (NREL) for conversations about stochastic optimization programs and Colin Campbell (Infleqtion) for conversations about probability distributions on quantum computers. CR would like to thank Eliot Kapit (Colorado School of Mines) for conversations about the adiabatic theorem and quantum annealing.

This work was authored by the National Renewable Energy Laboratory, managed and operated by Alliance for Sustainable Energy, LLC for the U.S. Department of Energy (DOE) under Contract No. DE-AC36-08G028308. 
This work was supported in part by the Laboratory Directed Research and Development (LDRD) Program at NREL. A portion of this research was performed using computational resources sponsored by the U.S. Department of Energy's Office of Energy Efficiency and Renewable Energy and located at the National Renewable Energy Laboratory. The views expressed in the article do not necessarily represent the views of the DOE or the U.S. Government. The U.S. Government retains and the publisher, by accepting the article for publication, acknowledges that the U.S. Government retains a nonexclusive, paid-up, irrevocable, worldwide license to publish or reproduce the published form of this work, or allow others to do so. for U.S. Government purposes.

\textbf{Author contributions} C.R. conceptualized the quantum algorithm along with its corresponding theory, wrote the code,  ran the experiments, and prepared the figures. M.R., P.G., and W.J. created the example problem. C.R., M.R., and W.J. formatted the problem in the context of nested solvers. C.R. and J.W. developed the annealing convergence argument as seen in Sec.~\ref{subsubsec:adiabatic_stable_state}. C.C., P.G. and C.R. made the arguments about probability distributions. C.R., M.R., and P.G. developed the error convergence argument as seen in Sec.~\ref{subsec:error}. C.R., M.R., J.W., E.B.J. and P.G. helped prepare the manuscript.

\appendix

\section{Residual temperature in the approximately optimized wavefunction}\label{appendix:residual_temp}

In this appendix, we give a more detailed form of the expression $\langle H_Q \rangle_\xi = \phi(x)+\delta$ from Eq.~\ref{eq:variational_error} in Sec.~\ref{subsec:pdfqaoa}, defined w.r.t. the approximately optimized wavefunction $\ket{\tilde{\psi}_x^*}$ from Eq.~\ref{eq:trail_qaoa}. Starting with the approximately optimized wavefunction
\begin{equation}
\begin{split}
    \ket{\tilde{\psi}_x^*} &= U(\theta,T)\ket{x}\ket{0}\ket{0} \\
    &= \ket{x}\left[ \sum_{\omega\in\Omega} \sqrt{p(\omega)}\left(\alpha_{\omega,y_\omega^*}\ket{y_\omega^*} + \sum_{y\neq y_\omega^*}\alpha_{\omega,y}\ket{y}\right)\ket{\xi_\omega}\right],
\end{split}
\end{equation}
where the $\alpha$ coefficients represent how converged each scenario is, and for each scenario $\omega$ we have $\sum_y |\alpha_{\omega,y}|^2 = 1$. We compute its expectation value $\langle H_Q \rangle_\xi$ as the following:
\begin{align}
    \langle H_Q \rangle_\xi &= \bra{\tilde{\psi}_x^*} H_Q \ket{\tilde{\psi}_x^*}\\
    &=\bra{\tilde{\psi}_x^*} \left(\sum_{\omega \in \Omega} H_{q_\omega} \otimes \ket{\xi_\omega}\bra{\xi_\omega}\right) \ket{\tilde{\psi}_x^*}.
\end{align}
By expanding, and combining overlapping basis vectors when possible, we get
\begin{equation}
\begin{split}
    \langle H_Q\rangle_\xi &= \bra{\tilde{\psi}_x^*} \left(\sum_{\omega \in \Omega} H_{q_\omega} \otimes \ket{\xi_\omega}\bra{\xi_\omega}\right) \ket{x}\left[ \sum_{\omega\in\Omega} \sqrt{p(\omega)}\left(\alpha_{\omega,y_\omega^*}\ket{y_\omega^*} + \sum_{y\neq y_\omega^*}\alpha_{\omega,y}\ket{y}\right)\ket{\xi_\omega}\right]\\
    &= \bra{\tilde{\psi}_x^*} \Bigg(\sum_{\omega \in \Omega} \sqrt{p(\omega)}H_{q_\omega}\ket{x}\left(\alpha_{\omega,y_\omega^*}\ket{y_\omega^*} + \sum_{y\neq y_\omega^*}\alpha_{\omega,y}\ket{y}\right)\ket{\xi_\omega} \Bigg)\\
    &= \bra{\tilde{\psi}_x^*}\Bigg(\sum_{\omega \in \Omega} \sqrt{p(\omega)}\Bigg(q(x,y_\omega^*,\xi_\omega)\alpha_{\omega,y_\omega^*}\ket{x}\ket{y_\omega^*} + \sum_{y\neq y_\omega^*}q(x,y,\xi_\omega)\alpha_{\omega,y}\ket{x}\ket{y}\Bigg)\ket{\xi_\omega} \Bigg),
\end{split}
\end{equation}
and if $\alpha^\dagger$ is the complex conjugate of $\alpha$,
\begin{equation}
\begin{split}
    \langle H_Q\rangle_\xi &= \bra{x}\left(\sum_{\omega\in\Omega}\sqrt{p(\omega)}\left(\alpha_{\omega,y_\omega^*}^\dagger \bra{y_\omega^*} + \sum_{y\neq y_\omega^*}\alpha_{\omega,y}^\dagger\bra{y} \right)\bra{\xi_\omega} \right) \\
    &\hspace{.6in}\left(\sum_{\omega \in \Omega} \sqrt{p(\omega)}\left(q(x,y_\omega^*,\xi_\omega)\alpha_{\omega,y_\omega^*}\ket{x}\ket{y_\omega^*} + \sum_{y\neq y_\omega^*}q(x,y,\xi_\omega)\alpha_{\omega,y}\ket{x}\ket{y}\right)\ket{\xi_\omega} \right)\\
    &= \sum_{\omega\in\Omega}p(\omega) \left(|\alpha_{\omega,y_\omega^*}|^2q(x,y_\omega^*,\xi_\omega) + \sum_{y\neq y_\omega^*} |\alpha_{\omega,y}|^2q(x,y,\xi_\omega) \right)\\
    &= \phi(x) + \sum_{\omega\in\Omega}p(\omega) \sum_{y\neq y_\omega^*}|\alpha_{\omega,y}|^2 q(x,y,\xi_\omega) - \sum_{\omega\in\Omega}p(\omega)\left(1-|\alpha_{\omega,y_\omega^*}|^2\right)q(x,y_\omega^*,\xi_\omega)\\
    &= \phi(x) + \sum_{\omega\in\Omega}p(\omega)\left[ \sum_{y\neq y_\omega^*}|\alpha_{\omega,y}|^2 q(x,y,\xi_\omega) + |\alpha_{\omega,y_\omega^*}|^2q(x,y_\omega^*,\xi_\omega) - q(x,y_\omega^*,\xi_\omega) \right].
\end{split}
\end{equation}
This gives us the following expression for the residual temperature:
\begin{equation}
    \delta = \sum_{\omega\in\Omega}p(\omega)\left[ \sum_{y\neq y_\omega^*}|\alpha_{\omega,y}|^2 q(x,y,\xi_\omega) + |\alpha_{\omega,y_\omega^*}|^2q(x,y_\omega^*,\xi_\omega) - q(x,y_\omega^*,\xi_\omega) \right].
\end{equation}
Recall that $\sum_y |\alpha_{\omega,y}|^2 = 1$, $\forall \omega \in \Omega$. Additionally, recall that $\lim_{T\rightarrow \infty}\alpha_{\omega,y_\omega^*} = 1$, $\forall \omega \in \Omega$. Therefore, we can see that as $T\rightarrow \infty $, we get $\delta \rightarrow 0$.

\section{Consistency of the oracle and normalization in the presence of residual temperature}\label{appendix:oracle_res_temp}
In this appendix we show that the oracle $\mathcal{F}$ used for QAE behaves as expected when the wavefunction still has residual temperature; in other words, that the probability of measuring a $\ket{1}$ in the ancilla qubit is $(\langle H_Q \rangle_\xi - q_l)/(q_u-q_l)$, and that this value is still bound on $[0,1]$. 

First, take the expression for $\ket{\tilde{\psi}_x^*}$ from Eq.~\ref{eq:trail_qaoa},
\begin{equation}
    \ket{\tilde{\psi}_x^*} = U(\theta,T)\ket{x}\ket{0} = \ket{x}\left[ \sum_{\omega\in\Omega} \sqrt{p(\omega)}\left(\alpha_{\omega,y_\omega^*}\ket{y_\omega^*} + \sum_{y\neq y_\omega^*}\alpha_{\omega,y}\ket{y}\right)\ket{\xi_\omega}\right],
\end{equation}
and the oracle $\mathcal{F}$ from Eq.~\ref{eq:oracle},
\begin{equation}
    \mathcal{F}\ket{x}\ket{y}\ket{\xi_\omega}\ket{0} = \ket{x}\ket{y}\ket{\xi_\omega}\left( \sqrt{1-\bar{q}(x,y,\xi_\omega)} \ket{0} + \sqrt{\bar{q}(x,y,\xi_\omega)}\ket{1}\right),
\end{equation}
where $\bar{q}(x,y,\xi_\omega) = (q(x,y,\xi_\omega) - q_l)/(q_u - q_l)$. Computing $\mathcal{F}\ket{\tilde{\psi}_x^*}$, we get
\begin{equation}
\begin{split}
    \mathcal{F}\ket{\tilde{\psi}_x^*}\ket{0} &= \mathcal{F}\ket{x}\left[ \sum_{\omega\in\Omega} \sqrt{p(\omega)}\left(\alpha_{\omega,y_\omega^*}\ket{y_\omega^*} + \sum_{y\neq y_\omega^*}\alpha_{\omega,y}\ket{y}\right)\ket{\xi_\omega}\right]\ket{0}\\
    &= \ket{x}\sum_{\omega\in\Omega} \sqrt{p(\omega)}\Bigg(\alpha_{\omega,y_\omega^*}\ket{y_\omega^*}\ket{\xi_\omega}\left(\sqrt{1-\bar{q}(x,y_\omega^*,\xi_\omega)}\ket{0} +\sqrt{\bar{q}(x,y_\omega^*,\xi_\omega)}\ket{1}\right)  \\
    &\hspace{.6in} + \sum_{y\neq y_\omega^*}\alpha_{\omega,y}\ket{y}\ket{\xi_\omega} \left(\sqrt{1-\bar{q}(x,y,\xi_\omega)}\ket{0} +\sqrt{\bar{q}(x,y,\xi_\omega)}\ket{1} \right)\Bigg).
\end{split}
\end{equation}
Immediately following this, we can see that the probability of measuring a $\ket{1}$ in the ancilla qubit is  
\begin{align}\label{eq:appendix:pr_ancilla}
\text{Pr}\left[\ket{1}\right] &= \sum_{\omega \in \Omega} p(\omega) \left( |\alpha_{\omega,y_\omega^*}|^2 \bar{q}(x,y_\omega^*,\xi_\omega) + \sum_{y\neq y_\omega^*} |\alpha_{\omega,y}|^2 \bar{q}(x,y,\xi_\omega) \right),
\end{align}
which we can combine with the bound $q\in[q_l,q_u]$ and the derivation in Appendix~\ref{appendix:residual_temp} to get
\begin{equation}
\begin{split}
   \text{Pr}\left[\ket{1}\right] &= \sum_{\omega\in\Omega}p(\omega) \left( |\alpha_{\omega,y_\omega^*}|^2 \frac{q(x,y_\omega^*,\xi_\omega)-q_l}{q_u-q_l} + \sum_{y\neq y_\omega^*} |\alpha_{\omega,y}|^2 \frac{q(x,y,\xi_\omega)-q_l}{q_u-q_l}  \right)\\ 
   &= \frac{1}{\left(q_u-q_l\right)} \left(\sum_{\omega\in\Omega}p(\omega) \left( |\alpha_{\omega,y_\omega^*}|^2 q(x,y_\omega^*,\xi_\omega) + \sum_{y\neq y_\omega^*} |\alpha_{\omega,y}|^2 q(x,y,\xi_\omega) \right) -q_l \right)\\
   &= \frac{\langle H_Q \rangle_\xi - q_l}{q_u - q_l}.
\end{split}
\end{equation}
Therefore, without changing the normalization, the QAE procedure will produce the estimated expected value function via $\text{Pr}[\ket{1}] = (\phi(x) + \delta - q_l )/( q_u - q_l)$ as explained in Sec.~\ref{sec:qae}. 

We also want to show the bound on the expected value function with residual temperature  $\phi(x) + \delta \in [q_l, q_u]$ holds. This will prove that the error bound given by Eq.~\ref{eq:error_bound} holds. Proving this is equivalent to proving that $0\leq (\phi(x) + \delta - q_l)/(q_u-q_l) \leq 1$. In other words:
\begin{equation}
    \phi(x) + \delta \in \left[q_l, q_u\right] \; \Leftrightarrow \; 0\leq \frac{\phi(x) + \delta - q_l}{q_u-q_l} \leq 1
\end{equation}
This bound may seem obvious because of the probability statement given above; nevertheless, we still show this to strengthen our argument. Starting with the form given in Eq.~\ref{eq:appendix:pr_ancilla}, and absorbing the optimal second stage decision $y_\omega^*$ into the sum (it is not needed to be marked as distinct for this calculation)
\begin{align}
    \frac{\phi(x)+\delta-q_l}{q_u-q_l} &= \sum_{\omega \in \Omega} p(\omega) \sum_{y} |\alpha_{\omega,y}|^2 \bar{q}(x,y,\xi_\omega)
\end{align}
additionally, recall that $1=\sum_{y}|\alpha_{\omega,y}|^2$ and $0\leq\bar{q}\leq 1$. If a sequence of positive fractions sums to one, and then each fraction in the sequence is multiplied by a positive fraction, the result must be bound on zero to one. Therefore, 
\begin{equation}
0\leq\sum_y |\alpha_{\omega,y}|^2\bar{q}(x,y,\xi_\omega)\leq 1.
\end{equation}
Again, $\sum_{\omega\in\Omega}p(\omega) = 1$, so we get 
\begin{equation}
\begin{split}
0 \leq \sum_{\omega \in \Omega} p(\omega)& \sum_{y} |\alpha_{\omega,y}|^2 \bar{q}(x,y,\xi_\omega) \leq 1\\
 &\Leftrightarrow \; 0\leq\frac{\phi(x)+\delta - q_l}{q_u - q_l}\leq 1.
\end{split}
\end{equation}
Because of the right inequality, we know that the expected value function plus residual temperature must be bound as $\phi(x)+\delta \in [q_l, q_u]$.

\printbibliography 

\end{document}